\documentclass[oneside]{article}
\usepackage[T1]{fontenc}
\usepackage{authblk}
\usepackage{float}
\usepackage{amsmath}
\usepackage{graphicx}
\usepackage{xfrac}
\usepackage[english]{babel}
\usepackage{lmodern}
\usepackage[T1]{fontenc}
\usepackage[latin9]{inputenc}

\usepackage{csquotes}

\usepackage{mathtools}
\usepackage{graphicx}
\usepackage{esint}
\usepackage[unicode=true,
 bookmarks=false,
breaklinks=false,pdfborder={0 0 1},backref=false,colorlinks=false]
 {hyperref}
\usepackage[a4paper, total={210mm,297mm},margin=2.0cm]{geometry}

\title{Elucidating the mechanism of step-emulsification} 

\author[1]{Andrea Montessori \thanks{Electronic address: \texttt{and.montessori@gmail.com}; Corresponding author}}

\author[1]{Marco Lauricella}

\author[2,1,3]{Sauro Succi}

\author[4]{Elad Stolovicki}

\author[4,5]{David Weitz}

\affil[1]{Istituto per le Applicazioni del Calcolo CNR, via dei Taurini 19, Rome, Italy}
\affil[2]{Center for Life Nano Science@La Sapienza, Istituto Italiano di Tecnologia, 00161 Roma, Italy}
\affil[3]{Institute for Applied Computational Science, Harvard John A. Paulson School of Engineering And Applied Sciences, Cambridge, MA 02138, United States}
\affil[4]{School of Engineering and Applied Sciences, Harvard University, McKay 517 Cambridge, MA 02138, USA}
\affil[5]{Department of Physics, and School of Engineering and Applied Sciences, Harvard University, Pierce 231, 29 Oxford St. Cambridge, MA 02138, USA}

\date{\today}

\begin{document}

\maketitle

\begin{abstract}
Three-dimensional, time-dependent direct simulations of step emulsification micro-devices
highlight two essential mechanisms for droplet formation: first, the onset of
an adverse pressure gradient driving a back-flow of the continuous phase from the external 
reservoir to the micro-channel. Second, the striction of the flowing jet which leads to its subsequent rupture.
It is also shown that such a rupture is delayed and eventually suppressed by increasing the flow speed
of the dispersed phase within the channel, due to the stabilising effect of dynamic pressure.  
This suggests a new criterion for dripping-jetting transition, based on local values of the Capillary
and Weber numbers.
\end{abstract}

Step emulsification (SE) has captured significant interest in the recent years as
a viable microfluidic technique for the controlled production of liquid droplets \cite{Sugiura2002,Priest2006}.
Among others, one of the main appeals of the SE technique
is the prospect of producing large volume rates of the dispersed phase, which are out
of reach for previous techniques, such as flow-focusers \cite{garstecki2005mechanism,garstecki2006formation,costantini2016correlation,anna2003formation}.

\begin{figure}[h!]
\centering
\includegraphics[width=0.9\columnwidth]{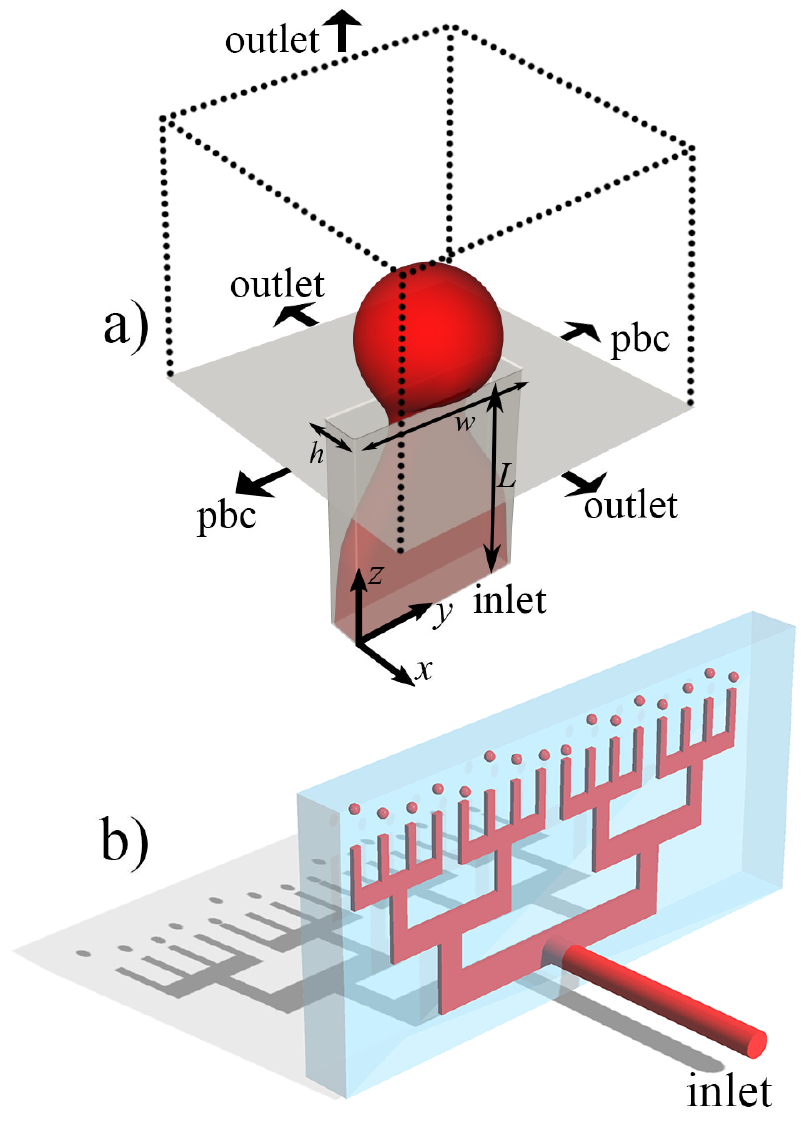}
\caption{Sketch of the nozzle geometry in the simulation box, along with the imposed boundary conditions 
(top panel a). The adopted conditions reproduce a periodic array of independent nozzles, which is consistent with the 
geometry of the volcano device (bottom panel b).
Here, the dispersed phase (red) is pumped through the device, forming mono-disperse drops in a 
reservoir containing a continuous immiscible phase (cyan).
}
\label{schema_finale}
\end{figure}

The basic idea behind SE is to exploit the pressure drop  due to
a sudden channel expansion (step) to induce the pinch-off of the  dispersed phase, leading to droplet formation \cite{dangla2013physical}. 
Albeit conceptually straightforward, the details of the process depend on a number of physical
and geometrical parameters, primary the capillary number $Ca$ and the aspect ratio
$h/w$ of the height versus width of the micro channel cross-section (see Figure \ref{schema_finale}).
Such parameters dictate the shape of the droplet and the transition between 
the  dripping and the jetting regimes\cite{dangla2013physical,stolovicki2017throughput,Ofner2017}.
Although of primary importance, the Capillary number (viscous dissipation/surface tension)
does not capture the full picture and needs to be complemented by other dimensionless
groups, namely the Weber number (inertia/surface tension) and/or Reynolds numbers (inertia/viscous dissipation).

Despite major technological advances, the theoretical description and the numerical simulation 
of micro-channel emulsification is still under development.

In this Letter,  we present direct numerical simulations of the fully three-dimensional, time-dependent
Navier-Stokes equations for a specific step-emulsification micro-device, in order to elucidate
the basic fluid phenomena underpinning the step-emulsification process.
The simulations highlight two essential mechanisms:  
i) the backflow of the continuous phase from the external reservoir to the
confined micro channel, driven by an adverse pressure gradient, 
ii) the resulting striction of the flowing jet within the channel and its subsequent rupture.
It is also shown that such a rupture is delayed or even suppressed upon increasing the flow speed
of the dispersed phase within the channel, due to the stabilising effect of dynamic pressure.

\begin{figure*}[t]
\centering+
\includegraphics[scale=0.8]{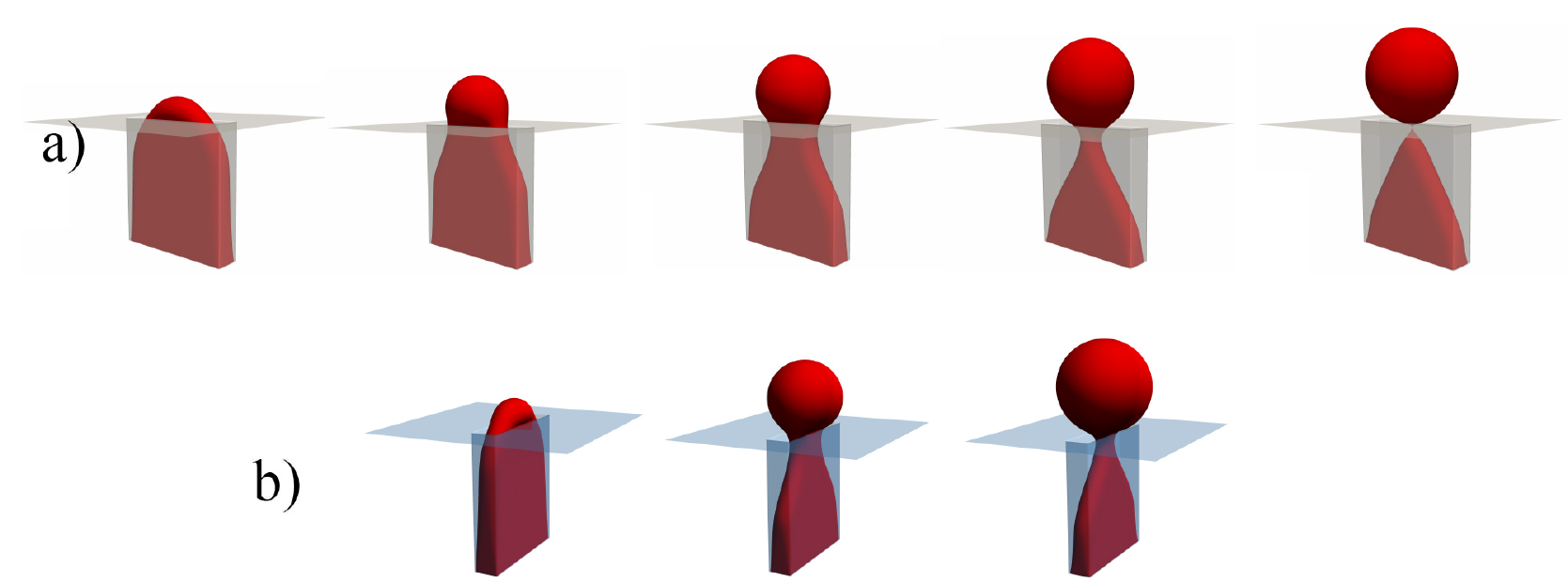}
\caption{\label{sketch} a) sequence of the break process in the dripping regime in a step emulsification nozzle. The dispersed phase (water), is 
pumped through the device and forms monodisperse drops whose sizes are proportional to
the nozzle height.
b) Sequence of the step emulsifier in the jetting regime}
\end{figure*}

In order to simulate the droplet breakup in a recently proposed  class of step emulsification devices \cite{Mittal2014,Priest2006,stolovicki2017throughput} we solve the multi-component Navier-Stokes equations, using an extensions  of the Lattice Boltzmann method \cite{MONTESSORI201833,higuera1989lattice,montessori2015lattice,benzi2009mesoscopic,succi2018lattice,leclaire2012numerical,reis2007lattice}.
See Supplemental Material at \ref{supp} for a detailed description of the numerical model employed
\cite{leclaire2011isotropic,montessori2014regularized,zhang2006efficient}.
The device, made of  polydimethylsiloxane (PDMS), is used  for producing water in oil emulsions and a sketch is reported in figure \ref{schema_finale}.
The water flows through the device inlet, and splits into hundreds of step-emulsifier nozzles with rectangular cross sections.
The PDMS device is submerged in quiescent oil, the continuous phase, with the
nozzles pointing upwards. The dispersed phase (water), is then
pumped through the device and forms monodisperse drops, whose sizes are 
proportional to the nozzle height ($h$).
We wish to point out that, being the device submerged in quiescent oil, there is  no net flow 
of the continuous phase, this in contrast with the emulsification system employed in \cite{li2015step}.
In this work, we simulate a single nozzle out of the full device, using periodic 
boundary conditions along crossflow directions, in order to mimic the effect of neighbouring nozzles.
At the inlet, we impose uniform velocity profiles, while a zero gradient approximation is applied at the outlet.
The ratio between the kinematic viscosities of the dispersed and the continuous phase is fixed to $\sim 1.2$, as in \cite{stolovicki2017throughput}, \footnote{We performed additional simulations in which the same range of capillary numbers was spanned by changing the viscosity $\nu$ and or the surface tension $\sigma$. The simulation results show that the transition from dripping to jetting  occurs almost  at the same critical Capillary number, corresponding to lower flow rates as the viscosity of the inner phase is increased.} \cite{Kobayashi2005,stoffel2012bubble,van2010effect,vladisavljevic2011effect}.

First, we perform a  comparison with the experiments reported in \cite{stolovicki2017throughput}, by investigating the effect of the main non-dimensional parameters i.e., the capillary number ($Ca=\rho U \nu /\sigma$) and the Weber number (We=$\rho U^2 L/  \sigma$), where $U$ is the  
velocity of the dispersed phase at the inlet, $\rho$ is the density of the dispersed phase, $L$ is a characteristic length defined as $\sqrt{wh}$ (see fig. \ref{schema_finale}), $\sigma$ is the oil-water surface tension and $\nu$ is the kinematic viscosity of the dispersed phase.

\begin{figure}
\centering
\includegraphics[scale=0.4]{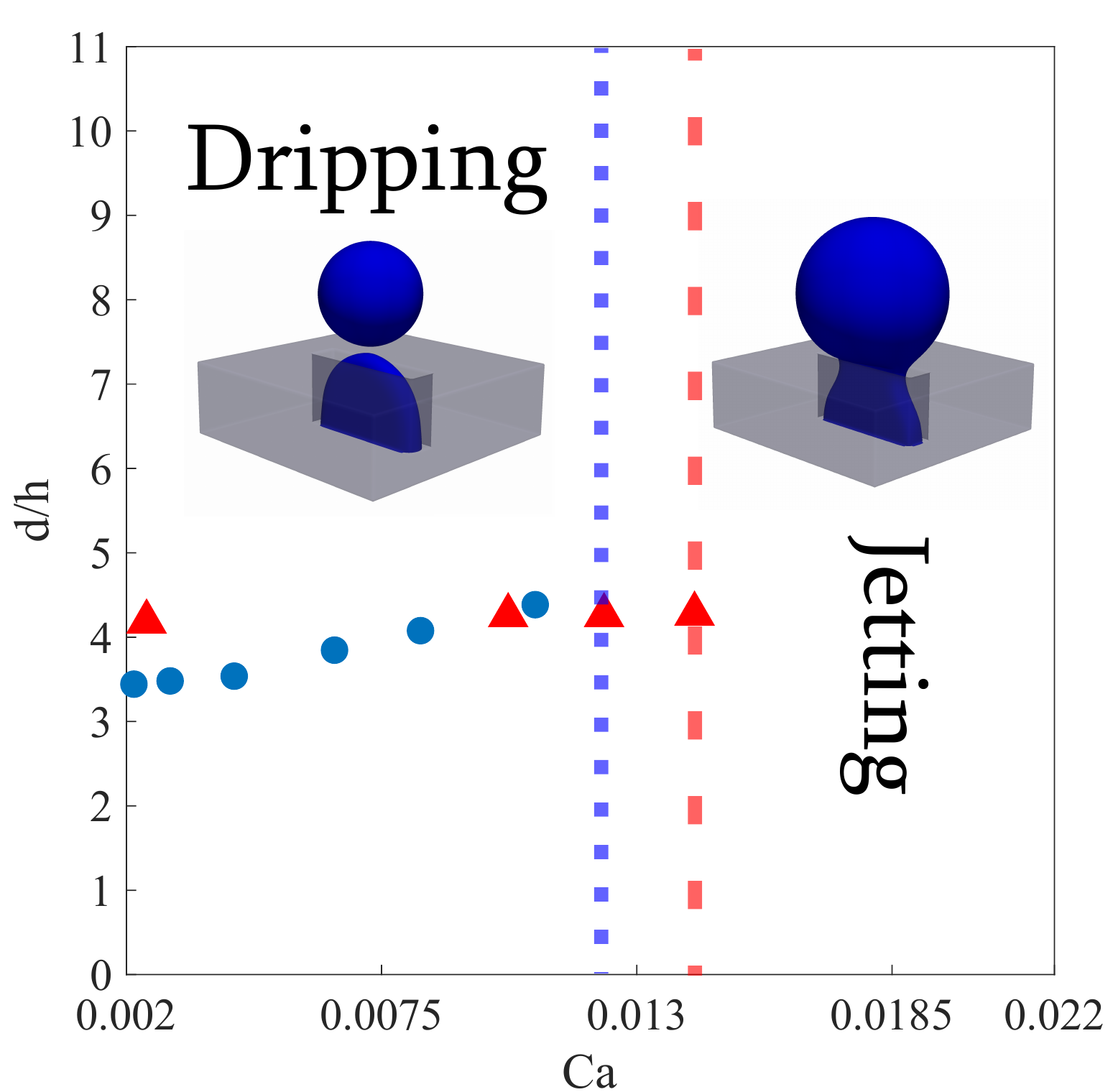}
\caption{\label{dovh} Dimensionless droplet diameter ($d/h$) versus the Capillary number for a nozzle aspect ratio $h/w=1/5$. The dots are the normalized droplet diameters predicted by the numerical simulations, while the  triangles stand for the experimental values of the normalized diameters.  In the dripping region, for $Ca$ between $ 0.002$ and $ \mathcal{O}(10^{-2})$ the average value of the droplet diameter is  $\sim 540 \mu m$ ($d/h \sim 4$).
The vertical dashed (experimental) and dotted (numerical) lines denote the critical Capillary numbers which mark the  dripping to jetting transition.  }
\end{figure}
In our simulations the dispersed phase flows through nozzles with a rectangular cross section of $135 \times 700 \mu m^2$ and length $L=810 \mu m$,  with a characteristic 
nozzle aspect ratio $h/w=1/5$, being $h$ and $w$ the height and the width of the microchannel, respectively. 
Experiments show that the droplet sizes are nearly  independent of the flow rate over an extended range (between $12$ and $70\; mL/min$, $0.002<Ca<0.014$) with an  average drop diameter of $d=567 \pm 6 \mu m$.
In this range of flow rates, thus for Capillary numbers in the range $\mathcal{O}(10^{-3}) \div \mathcal{O}(10^{-2})$, the step emulsifier has been shown to operate in the dripping regime, while
for larger flow rates, i.e. for Capillary numbers larger than a critical value, a transition from dripping to jetting regime occurs, which 
is characterized by the production of much larger and polydisperse droplets \cite{Priest2006, Mittal2014}.
The dripping-jetting transition occurs whenever  the droplet does not break up anymore and starts 'ballooning' (i.e., above a critical Capillary number the newborn droplet  grows larger and larger, see fig. \ref{sketch}.
The simulations exhibit a satisfactory agreement with the experimental results, as shown in fig \ref{dovh}.
First, in the dripping region ($ 0.002 <Ca<0.0125$) the average droplet diameter 
is  $d \sim 540 \mu m$ ($d/h \sim 4$), in satisfactory agreement with the experimental data \cite{stolovicki2017throughput}, see fig. \ref{dovh}.
\begin{figure*}
\centering
\includegraphics[scale=1.0]{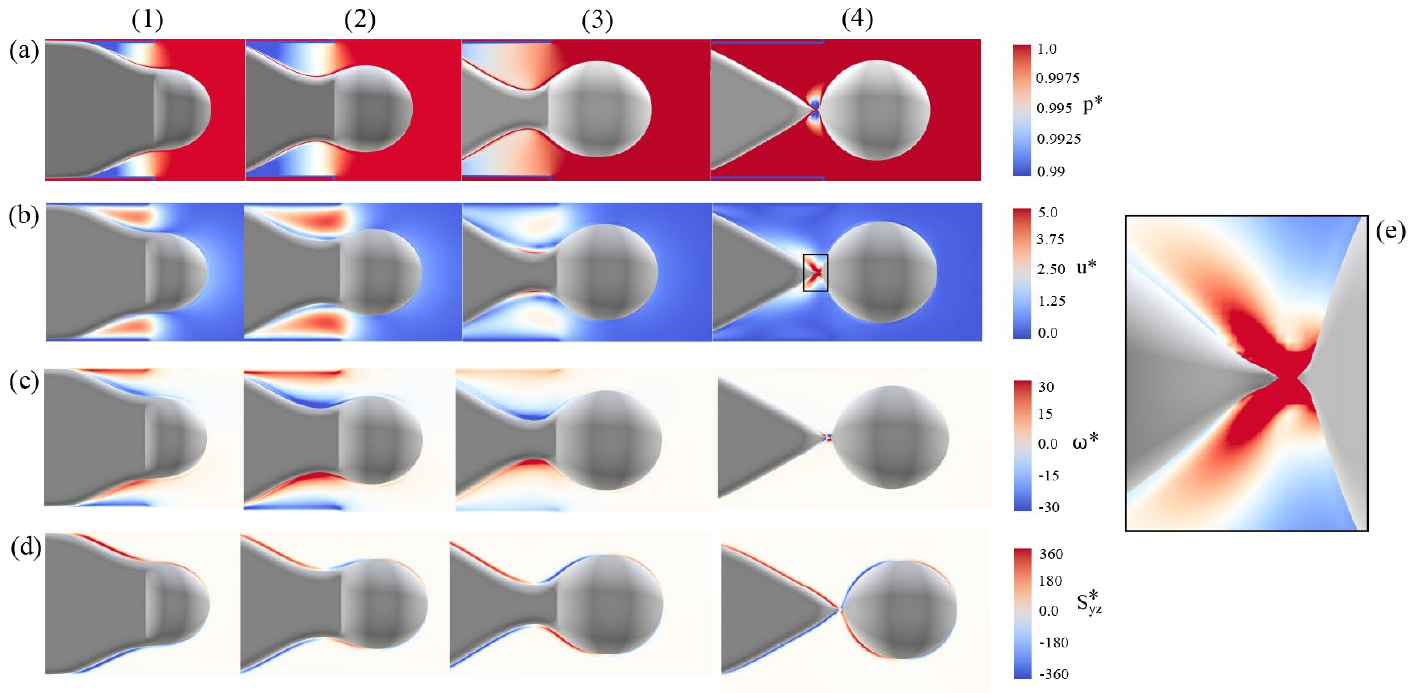}
\caption{\label{sequence}  (a) Pressure ($p^*$), (b) velocity ($u^*$), (c) vorticity ($\omega^*$) and (d) stress fields ($S^{*}_{yz}$), in a y-z  midplane taken between the two walls separated by $h$, from the focusing stage (1,2) to the pinching (3) and
finally breakup (4). (e) Zoom of the droplet structure and associated flow field at breakup time. In this simulation $Ca=0.003$ and $h/w=1/5$.}
\end{figure*}

The numerical simulations also predict a critical Capillary number $Ca \sim 0.0125$, which is in good agreement with  experimental observations \cite{stolovicki2017throughput,Mittal2014} (see fig. \ref{dovh}).
It is worth noting that, by scanning the capillary number, we encompass a broad
range of conditions associated with different  values of the physical properties,
such as dynamic viscosity, surface tension, as well as operational parameters, such as the flow rate.
 
Figure \ref{sequence} collects the main results of the present analysis.
In the top panel (a) we show a time-sequence of the pressure field from the focusing stage (1,2) to the pinching (3) and
finally breakup (4). This sequence unveils the following picture: 
the continuous phase flows back from the external reservoir to the
confined micro channel (focusing stage) and the flowing jet ruptures as a 
consequence of the striction induced by such backflow.
Note that the rupture is driven by the negative curvature which develops 
in the striction region (pinching stage).

In panel (b), we show the magnitude of the flow field, scaled with the inlet velocity. In figures b1 and b2, the build-up of a significant
back flow is apparent, amounting to about three times the inlet velocity. As the pinching progresses, the backflow speed decreases, due
to the enlarged section available to the continuum phase. At breakup time, a very localised burst is observed, signalling 
the rupture of the jet. In panel e) we zoom into the structure of the droplet and associated flow field at breakup time.
Such figure clearly shows the re-entrant motion of the jet, accompanied by the rapid acceleration of the newborn droplet.     

In panel (c) we show the time-sequence of the vorticity field, in units of $U/h$.
This sequence displays a typical elongational flow structure, especially in figure \ref{sequence}, panel c3, which stretches the jet until rupture. 

Finally, in panel (d) we show the component of the stress field in the $yz$ mid-plane,  in units of $U/h$.
The sequence shows that the stress field is highly localised around the oil-water interface, with a 
null-point right at the pinch position \cite{Eggers1993}.   The present analysis is in line with previous studies \cite{van2009flows}, in which the breakup is not interpreted as  
due to a Plateau-Rayleigh instability \cite{Eggers2008}, but rather to the back flow of the continuum phase, triggered by the 
adverse pressure gradient which arises in correspondence with the focusing of the water jet.
We wish to emphasize that our analysis is fully dynamic and three-dimensional, i.e  it does not rely on 
any quasi-static assumption \cite{dangla2013physical}, nor on any axial-symmetry of the flow \cite{chakraborty2017microfluidic}.

\begin{figure*}
\centering
\includegraphics[scale=0.70]{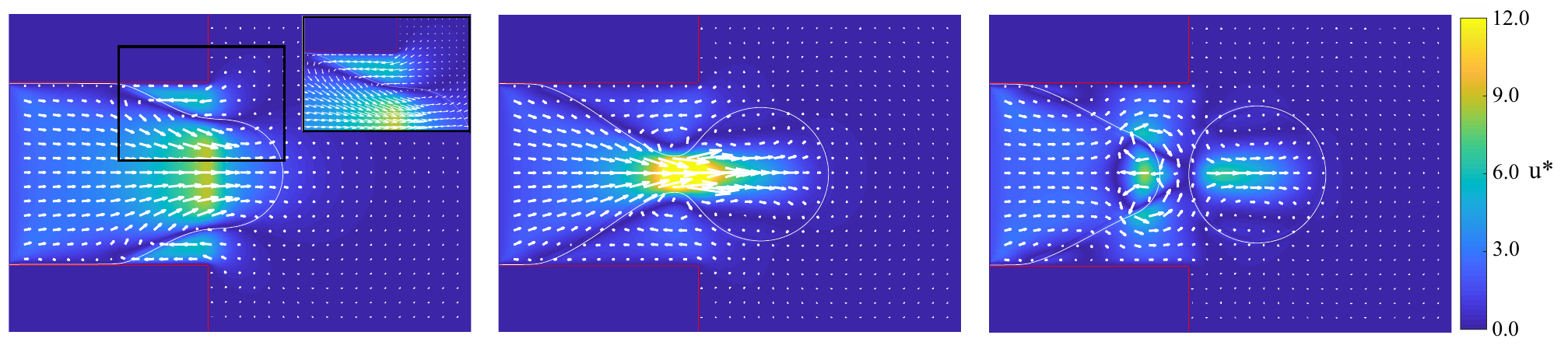}
\caption{\label{vpinch} Normalized velocity magnitude and vector field in the dripping nozzle. The counterflow in the continuous phase within the nozzle, is clearly evidenced by the quiver plot. The solid white line identifies the interface between the continuous and the dispersed phase while the red line denotes the walls of the nozzle. The inset in the leftmost panel, highlights  the backflow occurring inside the nozzle. }
\end{figure*}

 \textit{Dripping to jetting transition.}
 
Most experimental studies of step-emulsification report a dripping to jetting 
transition above a critical capillary number $Ca_{crit}\sim \mathcal{O}(10^{-2})$ \cite{Sugiura2002, stolovicki2017throughput, Mittal2014}. 
However, the underlying mechanisms behind such transition are still under investigation.
Here, we wish to point out that the transition to the jetting regime is facilitated by the contribution
of the dynamic pressure $\rho_{in} u_{in}^2/2$,being $\rho_{in}$ and $u_{in}$ the local density and velocity of the dispersed phase in the pinching region.
Such dynamic pressure  withstands the effects of 
the negative-curvature in the pinch region.
Due to the pinching effect, the local flow speed within the dispersed phase significantly exceeds the inlet velocity. Hence,
the local capillary number attains  larger values, of the order of $0.1-1$, whenever the nominal capillary 
number of the dispersed phase reaches its critical value around $0.01$. 
Note that the nominal capillary number is computed with the imposed velocity at the inlet.\\

The local acceleration of the flow field inside the pinch region is clearly visible in Figure \ref{vpinch}, which 
reports the flow field inside the neck region of the dispersed phase.

\begin{figure}
\includegraphics[scale=0.6]{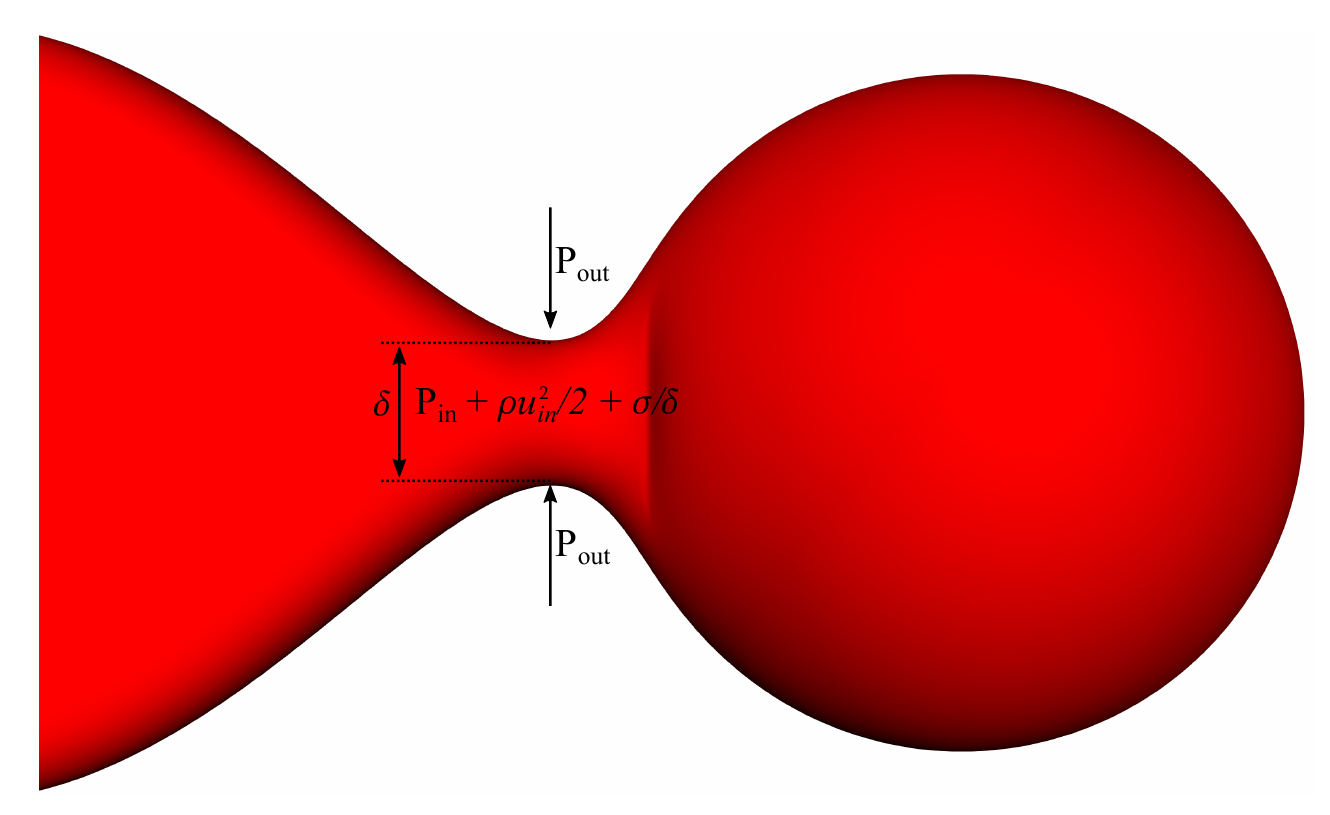}
\caption{\label{sketch_pressure} 
 Sketch of the static and dynamic pressure components in and out of the neck region.}
\end{figure}
As pinching progresses in time, the flow speed inside the pinching region grows accordingly, so that, at some point, inertial
effects can no longer be neglected.

To clarify the point, let us write the dynamic force balance under flow conditions, namely:
\begin{equation}\label{eq1}
P_{in} + \rho_{in} u_{in}^2/2= P_{out} + \rho_{out} u_{out}^2/2 - \frac{\sigma}{\delta} 
\end{equation}
where subscripts "in" and "out" refer to the neck region (see fig.\ref{sketch_pressure}), inside and outside the water jet, respectively.
Note that the minus sign in front of the surface tension reflects the negative curvature (see fig. \ref{sketch_pressure}).

In equation \ref{eq1}, $\delta$ is the characteristic length scale of the neck region, 
which is found to be smaller but comparable with the channel height $h$. 
This is plausible, because $\delta > h$ is not feasible since the neck diameter cannot be larger than the 
height of the nozzle, while $\delta \ll h$ signals the imminent breakup. 

This expression shows that the inner dynamic pressure adds to the surface tension in withstanding the
outer pressure. As a result, it is natural to extend the definition of Capillary number so as to 
include the contribution of the dynamic pressure, namely:
\begin{equation}
K = \frac{\mu u_{in}}{\sigma + \rho u_{in}^2 \delta/2}  \equiv \frac{Ca_{in}}{1+ We_{in}}
\end{equation}
where  we have neglected the outer velocity since $u_{out}/u_{in} \sim \delta/w < 1$.

For the case of figure  \ref{sequence}, we have $Ca_{in} = 0.015$ and $We_{in} \sim 0.22$, showing that the dynamic
pressure is still sub-dominant with respect to the capillary pressure $\sigma/\delta$.
On the other hand, in the case of jetting, (see panel (c) in Fig. \ref{sketch}) $We_{in} \sim 1$, indicating
that the jetting regime is entered whenever the dynamic pressure becomes comparable 
or higher than the capillary pressure. 

We wish to point out that since $\delta \sim h$, $Ca_{in} \sim \frac{w}{h}Ca $, which is precisely the
quantity controlling the dripping-jetting criterion discussed in \cite{li2015step}.
In this Letter, we noted that such criterion should also take into account the contribution
of dynamic pressure, which becomes dominant in the pinch region.    



\section{Acknowledgments}
The research leading to these results has received
funding from the European Research Council under the European
Union's Horizon 2020 Framework Programme (No. FP/2014-
2020)/ERC Grant Agreement No. 739964  (\textquotedbl{}COPMAT\textquotedbl{}).

\section{Supplementary information\label{supp}}
The LB immiscible multicomponent model is based on the following lattice Bhatnagar-Gross-Krook (BGK) equation:
\begin{equation}
f_{i}^{k} \left(\vec{x}+\vec{c}_{i}\Delta t,\,t+\Delta t\right) =f_{i}^{k}\left(\vec{x},\,t\right)+\Omega_{i}^{k}( f_{i}^{k}\left(\vec{x},\,t\right)),
\end{equation}
where $f_{i}^{k}$ is the discrete distribution function, representing
the probability of finding a particle of the $k-th$ component at position $\vec{x}$ and time
$t$ with discrete velocity $\vec{c}_{i}$ . 
The lattice time step is taken equal to 1, and $i$ the index spans the lattice discrete directions $i = 0,...,b$, where $b=26$ for a two dimensional nine speed lattice (D3Q27).
The density $\rho^{k}$ of the $k-th$ fluid component is given by the zeroth order moment of the distribution functions
\begin{equation}
\rho^{k}\left(\vec{x},\,t\right) = \sum_i f_{i}^{k}\left(\vec{x},\,t\right),
\end{equation}
while the total momentum $\rho \vec{u} $ is defined by the first order moment:
\begin{equation}
\rho \vec{u} = \sum_i  \sum_k f_{i}^{k}\left(\vec{x},\,t\right) \vec{c}_{i}.
\end{equation}
The collision operator $\Omega_{i}^{k}$ results from the combination of three sub-operators, namely \cite{leclaire2012numerical}
\begin{equation}
\Omega_{i}^{k} = \left(\Omega_{i}^{k}\right)^{(3)}\left[\left(\Omega_{i}^{k}\right)^{(1)}+\left(\Omega_{i}^{k}\right)^{(2)}\right].
\end{equation}
Here, $\left(\Omega_{i}^{k}\right)^{(1)}$ is the standard BGK operator for the $k-th$
component, accounting for relaxation towards a local equilibrium
\begin{equation}
\left(\Omega_{i}^{k}\right)^{(1)} f_{i}^{k}\left(\vec{x},\,t\right) = f_{i}^{k}\left(\vec{x},\,t\right) - \omega_{k}\left(f_{i}^{k}\left(\vec{x},\,t\right)-f_{i}^{k,eq}\left(\vec{x},\,t\right)\right),
\end{equation}
where $\omega_{k}$ is the relaxation rate, and $f_{i}^{k,eq}\left(\vec{x},\,t\right)$ denotes local equilibria, defined by
\begin{equation}
f_{i}^{k,eq}\left(\vec{x},\,t\right)=  \rho^{k} w_i \left( 1 +  \frac{\vec{c}_i \cdot \vec{u}}{c_s^2}  + \frac{(\vec{c}_i \cdot \vec{u} )^2}{2c_s^4} -  \frac{( \vec{u} )^2}{2c_s^2} \right) 
\end{equation}
Here, $w_i$ are weights of the discrete equilibrium distribution functions, 
$c_s=1/\sqrt(3)$ is the lattice sound speed \cite{succi2001lattice}.
In this model, $\left(\Omega_{i}^{k}\right)^{(2)}$ is a perturbation operator, modeling the surface tension of the $k-th$ component.
Denoting by $\vec{F}$ the color gradient in terms of the color difference (see below), this term reads
\begin{equation}
\left(\Omega_{i}^{k}\right)^{(2)} f_{i}^{k}\left(\vec{x},\,t\right)= f_{i}^{k}\left(\vec{x},\,t\right)+A_k|\vec{F}| \left[w_i \frac{(\vec{F} \cdot \vec{c}_i)^2}{|\vec{F}|^2} -B_i \right],
\end{equation}
with the free parameters $A_k$ modeling the surface tension, and $B_k$ a parameter depending on the
chosen lattice \cite{reis2007lattice,leclaire2011isotropic}.
The above operator models the surface tension, but it does not guarantee the immiscibility between different components. 
In order to minimize the mixing of the fluids, a recoloring operator $\left(\Omega_{i}^{k}\right)^{(3)}$  is introduced.
Following the approach in Ref. \cite{leclaire2011isotropic}, being  $\zeta$ and $\xi$ two immiscible fluids, the recoloring operators 
for the two fluids read as follows
\begin{equation}
\begin{aligned}
\left(\Omega_{i}^{\zeta}\right)^{(3)} &= \frac{\rho^{\zeta}}{\rho}f_i + \beta \frac{\rho^{\zeta} \rho^{\xi}}{\rho^2} \cos(\phi_i) \sum_{k=\zeta,\xi} f_i^{k,eq} (\rho^k , 0) \\
\left(\Omega_{i}^{\xi}\right)^{(3)} &= \frac{\rho^{\xi}}{\rho}f_i - \beta \frac{\rho^{\zeta} \rho^{\xi}}{\rho^2} \cos(\phi_i)\sum_{k=\zeta,\xi} f_i^{k,eq} (\rho^k , 0)
\end{aligned}
\end{equation}
where $\beta$ is a free parameter and $\cos(\phi_i)$ is the cosine of the angle between the color gradient $\vec{F}$
and the lattice direction $\vec{c}_i$.
Note that $f_i^{k,eq} (\rho^k , 0)$ stands for the set of equilibrium distributions of $k-th$ fluid
evaluated setting the macroscopic velocity to zero.
In the above equation, $f_i=\sum_k f_i^k$.
The LB color gradient model has been enriched with the so called regularization procedure \cite{montessori2014regularized,zhang2006efficient,falcucci2017effects}, namely a discrete Hermite projection of the post-collisional set of distribution functions onto a proper set of Hermite basis.
The main idea is to introduce a set of pre-collision distribution functions which are defined only in terms of the 
macroscopic hydrodynamic moments.All the higher-order non-equilibrium information, often referred to as {\it ghosts} \cite{higuera1989lattice}, is discarded.In equations, the regularized LB reads as follows:
\begin{equation}
f_i^k(x_i+c_i \Delta t,t+\Delta t) = \mathcal{R} f_i^{k,neq}(x,t) \equiv f_i^{k,eq} -  \Delta t \omega_k  (f^{k,reg}_{i} - f_i^{k,eq} )
\end{equation}
where $f^{k,reg}_{i}$ is the hydrodynamic component of the full distribution $f_i^k$ (see \cite{zhang2006efficient}) for the $k-th$ color, and $\mathcal{R}$ 
is the regularization operator. The above equation shows that the post-collision distribution, of a $4^{th}$-order isotropic lattice, 
is defined only in terms of the conserved and the transport hydrodynamic modes, 
namely density $\rho$, current $\rho \vec{u}$ and momentum-flux tensor $\mathbf{\Pi}=\sum_i f_i \vec{c_i}\vec{c_i}$.


\begin{thebibliography}{10}
\expandafter\ifx\csname url\endcsname\relax
  \def\url#1{\texttt{#1}}\fi
\expandafter\ifx\csname urlprefix\endcsname\relax\def\urlprefix{URL }\fi
\expandafter\ifx\csname href\endcsname\relax
  \def\href#1#2{#2} \def\path#1{#1}\fi

\bibitem{Sugiura2002}
S.~Sugiura, M.~Nakajima, N.~Kumazawa, S.~Iwamoto, M.~Seki,
  \href{https://doi.org/10.1021/jp0259871}{Characterization of spontaneous
  transformation-based droplet formation during microchannel emulsification},
  The Journal of Physical Chemistry B 106~(36) (2002) 9405--9409.
\newblock \href {https://doi.org/10.1021/jp0259871}
  {\path{doi:10.1021/jp0259871}}.
\newline\urlprefix\url{https://doi.org/10.1021/jp0259871}

\bibitem{Priest2006}
C.~Priest, S.~Herminghaus, R.~Seemann,
  \href{https://doi.org/10.1063/1.2164393}{Generation of monodisperse gel
  emulsions in a microfluidic device}, Applied Physics Letters 88~(2) (2006)
  024106.
\newblock \href {https://doi.org/10.1063/1.2164393}
  {\path{doi:10.1063/1.2164393}}.
\newline\urlprefix\url{https://doi.org/10.1063/1.2164393}

\bibitem{garstecki2005mechanism}
P.~Garstecki, H.~A. Stone, G.~M. Whitesides, Mechanism for flow-rate controlled
  breakup in confined geometries: A route to monodisperse emulsions, Physical
  review letters 94~(16) (2005) 164501.

\bibitem{garstecki2006formation}
P.~Garstecki, M.~J. Fuerstman, H.~A. Stone, G.~M. Whitesides, Formation of
  droplets and bubbles in a microfluidic t-junction-scaling and mechanism of
  break-up, Lab on a Chip 6~(3) (2006) 437--446.

\bibitem{costantini2016correlation}
M.~Costantini, C.~Colosi, P.~Mozetic, J.~Jaroszewicz, A.~Tosato, A.~Rainer,
  M.~Trombetta, W.~{\'S}wi{\k{e}}szkowski, M.~Dentini, A.~Barbetta, Correlation
  between porous texture and cell seeding efficiency of gas foaming and
  microfluidic foaming scaffolds, Materials Science and Engineering: C 62
  (2016) 668--677.

\bibitem{anna2003formation}
S.~L. Anna, N.~Bontoux, H.~A. Stone, Formation of dispersions using "flow
  focusing" in microchannels, Applied physics letters 82~(3) (2003) 364--366.

\bibitem{dangla2013physical}
R.~Dangla, E.~Fradet, Y.~Lopez, C.~N. Baroud, The physical mechanisms of step
  emulsification, Journal of Physics D: Applied Physics 46~(11) (2013) 114003.

\bibitem{stolovicki2017throughput}
E.~Stolovicki, R.~Ziblat, D.~A. Weitz,
  \href{http://dx.doi.org/10.1039/C7LC01037K}{Throughput enhancement of
  parallel step emulsifier devices by shear-free and efficient nozzle
  clearance}, Lab on a Chip (2018).
\newblock \href {https://doi.org/10.1039/c7lc01037k}
  {\path{doi:10.1039/c7lc01037k}}.
\newline\urlprefix\url{http://dx.doi.org/10.1039/C7LC01037K}

\bibitem{Ofner2017}
A.~Ofner, D.~G. Moore, P.~A. Rühs, P.~Schwendimann, M.~Eggersdorfer,
  E.~Amstad, D.~A. Weitz, A.~R. Studart,
  \href{http://dx.doi.org/10.1002/macp.201600472}{High-throughput step
  emulsification for the production of functional materials using a glass
  microfluidic device}, Macromolecular Chemistry and Physics 218~(2) (2017)
  1600472--n/a, 1600472.
\newblock \href {https://doi.org/10.1002/macp.201600472}
  {\path{doi:10.1002/macp.201600472}}.
\newline\urlprefix\url{http://dx.doi.org/10.1002/macp.201600472}

\bibitem{Mittal2014}
N.~Mittal, C.~Cohen, J.~Bibette, N.~Bremond,
  \href{https://doi.org/10.1063/1.4892949}{Dynamics of step-emulsification:
  From a single to a collection of emulsion droplet generators}, Physics of
  Fluids 26~(8) (2014) 082109.
\newblock \href {https://doi.org/10.1063/1.4892949}
  {\path{doi:10.1063/1.4892949}}.
\newline\urlprefix\url{https://doi.org/10.1063/1.4892949}

\bibitem{MONTESSORI201833}
A.~Montessori, M.~Lauricella, M.~La~Rocca, S.~Succi, E.~Stolovicki, R.~Ziblat,
  D.~Weitz,
  \href{http://www.sciencedirect.com/science/article/pii/S0045793018300926}{Regularized
  lattice boltzmann multicomponent models for low capillary and reynolds
  microfluidics flows}, Computers \& Fluids 167 (2018) 33 -- 39.
\newblock \href
  {https://doi.org/https://doi.org/10.1016/j.compfluid.2018.02.029}
  {\path{doi:https://doi.org/10.1016/j.compfluid.2018.02.029}}.
\newline\urlprefix\url{http://www.sciencedirect.com/science/article/pii/S0045793018300926}

\bibitem{higuera1989lattice}
F.~Higuera, S.~Succi, R.~Benzi, Lattice gas dynamics with enhanced collisions,
  EPL (Europhysics Letters) 9~(4) (1989) 345.

\bibitem{montessori2015lattice}
A.~Montessori, P.~Prestininzi, M.~La~Rocca, S.~Succi, Lattice boltzmann
  approach for complex nonequilibrium flows, Physical Review E 92~(4) (2015)
  043308.

\bibitem{benzi2009mesoscopic}
R.~Benzi, M.~Sbragaglia, S.~Succi, M.~Bernaschi, S.~Chibbaro, Mesoscopic
  lattice boltzmann modeling of soft-glassy systems: theory and simulations,
  The Journal of Chemical Physics 131~(10) (2009) 104903.

\bibitem{succi2018lattice}
S.~Succi, The Lattice Boltzmann Equation: For Complex States of Flowing Matter,
  Oxford University Press, 2018.

\bibitem{leclaire2012numerical}
S.~Leclaire, M.~Reggio, J.-Y. Tr{\'e}panier, Numerical evaluation of two
  recoloring operators for an immiscible two-phase flow lattice boltzmann
  model, Applied Mathematical Modelling 36~(5) (2012) 2237--2252.

\bibitem{reis2007lattice}
T.~Reis, T.~Phillips, Lattice boltzmann model for simulating immiscible
  two-phase flows, Journal of Physics A: Mathematical and Theoretical 40~(14)
  (2007) 4033.

\bibitem{leclaire2011isotropic}
S.~Leclaire, M.~Reggio, J.-Y. Tr{\'e}panier, Isotropic color gradient for
  simulating very high-density ratios with a two-phase flow lattice boltzmann
  model, Computers \& Fluids 48~(1) (2011) 98--112.

\bibitem{montessori2014regularized}
A.~Montessori, G.~Falcucci, P.~Prestininzi, M.~La~Rocca, S.~Succi, Regularized
  lattice bhatnagar-gross-krook model for two-and three-dimensional cavity flow
  simulations, Physical Review E 89~(5) (2014) 053317.

\bibitem{zhang2006efficient}
R.~Zhang, X.~Shan, H.~Chen, Efficient kinetic method for fluid simulation
  beyond the navier-stokes equation, Physical Review E 74~(4) (2006) 046703.

\bibitem{li2015step}
Z.~Li, A.~Leshansky, L.~Pismen, P.~Tabeling, Step-emulsification in a
  microfluidic device, Lab on a Chip 15~(4) (2015) 1023--1031.

\bibitem{Kobayashi2005}
I.~Kobayashi, S.~Mukataka, M.~Nakajima, Effects of type and physical properties
  of oil phase on oil-in-water emulsion droplet formation in straight-through
  microchannel emulsification, experimental and cfd studies, Langmuir 21~(13)
  (2005) 5722--5730.

\bibitem{stoffel2012bubble}
M.~Stoffel, S.~Wahl, E.~Lorenceau, R.~H{\"o}hler, B.~Mercier, D.~E. Angelescu,
  Bubble production mechanism in a microfluidic foam generator, Physical review
  letters 108~(19) (2012) 198302.

\bibitem{van2010effect}
K.~van Dijke, I.~Kobayashi, K.~Schro{\"e}n, K.~Uemura, M.~Nakajima, R.~Boom,
  Effect of viscosities of dispersed and continuous phases in microchannel
  oil-in-water emulsification, Microfluidics and nanofluidics 9~(1) (2010)
  77--85.

\bibitem{vladisavljevic2011effect}
G.~T. Vladisavljevi{\'c}, I.~Kobayashi, M.~Nakajima, Effect of dispersed phase
  viscosity on maximum droplet generation frequency in microchannel
  emulsification using asymmetric straight-through channels, Microfluidics and
  Nanofluidics 10~(6) (2011) 1199--1209.

\bibitem{Eggers1993}
J.~Eggers,
  \href{https://link.aps.org/doi/10.1103/PhysRevLett.71.3458}{Universal
  pinching of 3d axisymmetric free-surface flow}, Phys. Rev. Lett. 71 (1993)
  3458--3460.
\newblock \href {https://doi.org/10.1103/PhysRevLett.71.3458}
  {\path{doi:10.1103/PhysRevLett.71.3458}}.
\newline\urlprefix\url{https://link.aps.org/doi/10.1103/PhysRevLett.71.3458}

\bibitem{van2009flows}
V.~van Steijn, C.~R. Kleijn, M.~T. Kreutzer, Flows around confined bubbles and
  their importance in triggering pinch-off, Physical review letters 103~(21)
  (2009) 214501.

\bibitem{Eggers2008}
J.~Eggers, E.~Villermaux,
  \href{http://stacks.iop.org/0034-4885/71/i=3/a=036601}{Physics of liquid
  jets}, Reports on Progress in Physics 71~(3) (2008) 036601.
\newline\urlprefix\url{http://stacks.iop.org/0034-4885/71/i=3/a=036601}

\bibitem{chakraborty2017microfluidic}
I.~Chakraborty, J.~Ricouvier, P.~Yazhgur, P.~Tabeling, A.~Leshansky,
  Microfluidic step-emulsification in axisymmetric geometry, Lab on a Chip
  17~(21) (2017) 3609--3620.

\end{thebibliography}

\end{document}